\documentclass[11pt]{revtex4-1}

\usepackage{graphicx}

\begin{document}

\title{IRREVERSIBLE THERMODYNAMICS OF THE UNIVERSE: CONSTRAINTS FROM PLANCK DATA}

\author{Subhajit Saha\footnote {subhajit1729@gmail.com}}
\affiliation{Department of Mathematics, Jadavpur University, Kolkata 700032, West Bengal, India}

\author{Atreyee Biswas\footnote {atreyee11@gmail.com}}
\affiliation{Department of Natural Sciences, West Bengal University of Technology, BF-142, Sector-1, Saltlake, Kolkata 700064, West Bengal, India}

\author{Subenoy Chakraborty\footnote {schakraborty.math@gmail.com}}
\affiliation{Department of Mathematics, Jadavpur University, Kolkata 700032, West Bengal, India.}

\begin{abstract}

The present work deals with irreversible Universal thermodynamics. The homogenous and isotropic flat model of the universe is chosen as open thermodynamical system and non-equilibrium thermodynamics comes into picture due to the mechanism of particle creation. For simplicity, entropy flow is considered only due to heat conduction. Further, due to Maxwell-Cattaneo modified Fourier law for non-equilibrium phenomenon, the temperature satisfies damped wave equation instead of heat conduction equation. Validity of generalized second law of thermodynamics (GSLT) has been investigated for Universe bounded by apparent or event horizon with cosmic substrutum as perfect fluid with constant or variable equation of state or interacting dark species. Finally, we have used three {\it Planck} data sets to constrain the thermal conductivity $\lambda$ and the coupling parameter $b^2$. These constraints must be satisfied in order for GSLT to hold for Universe bounded by apparent or event horizons.

\end{abstract}

\maketitle

\section{INTRODUCTION}

This is now well established that there is a profound relation between gravity and thermodynamics. In 1970's Hawking [1] and Bekenstein [2] gave rise to this unique idea with their revolutionary discovery of black hole thermodynamics. According to them, black hole behaves as a black body whose temperature (known as Hawking temperature) and entropy (known as Bekenstein entropy) are proportional to the surface gravity at the horizon and area of the horizon respectively. Later Bardeen {\it et al.} [3], in 1973, established that the four laws of black hole mechanics are actually analogous to four laws of thermodynamics. As thermodynamical parameters such as temperature and entropy are characterized by the geometry of the event horizon of the black hole, so it is legitimated to assume that black hole thermodynamics is deeply related to Einstein's field equations. This assertion became true when Jacobson [4] in 1995 succesfully derived Einstein equation from the first law of thermodynamics, $\delta Q=T dS$ with $\delta Q$ and $T$ as the energy flux and Unruh temperature measured by an accelerated observer just inside the horizon and subsequently Padmanabhan [5] derived the first law of thermodynamics from Einstein equations for general static spherically symmetric space time. Since then, much work have been done based on this equivalence between Einstein's equations and thermodynamics.

Universal thermodynamics got a new direction when it was understood that the Universe should be an irreversible one rather than a reversible one [6]. Jacobson [4] first noticed this when his attempt failed to reproduce the Einstein's equations from first law of thermodynamics in $f(R)$ gravity. In that case he assumed the horizon entropy to be proportional to a function of the Ricci scalar and this led to the break down of the local thermodynamical equilibrium. Subsequently, Eling {\it et al.} [6] had shown that by a curvature correction to the entropy which is polynomial in the Ricci scalar, Einstein's equations can be derived from thermodynamic laws in $f(R)$ gravity but it requires a non equilibrium treatment. In order to do so they added an extra term $d_{i}S$ called entropy production term  to the entropy balance equation,
\begin{center}
$dS=\frac{dQ}{T}+d_{i}S$,
\end{center} 
where they explained $d_{i}S$ as bulk viscosity production term determined by imposing energy-momentum conservation. In general, the entropy balance relation in non-equilibrium thermodynamics is of the form
\begin{center} 
$dS=d_{e}S+d_{i}S$,
\end{center} 
where $d_{e}S$ is the rate of entropy exchange with the surroundings while  $d_{i}S$ ($\geq 0$) comes from the process occuring inside the system. In particular  $d_{i}S$ is zero for reversible process and positive for irreversible process. In cosmology, $d_{i}S$ has no clear interpretation as it depends on the internal production process.

Gang {\it et al.} [7] studied non-equilibrium thermodynamics for Universe bounded by apparent horizon with dark energy in the form of perfect fluid with constant equation of state. They got an interesting result that the original radius of apparent horizon needs to be corrected and the new position of apparent horizon depends on constant equation of state of the dark energy as well as on the non-equilibrium factor.

In this paper, we have followed the work of Gang {\it et al.}. In particular, it is an extension of our previous works on non-equilibrium thermodynamics of Universe bounded by event [8] and apparent horizon [9]. In the next section, we have given a general description of irreversible process of the Universe. Section III deals with Universe bounded by apparent/event horizon for flat FRW model and cosmic substratum is chosen as the following three types:\\ 
a) Perfect fluid with constant equation of state,\\
b) perfect fluid with variable equation of state, and \\
c) interacting dark matter and holographic dark energy.\\
For each of the fluids validity of generalized second law of thermodynamics has also been examined for both the horizons. In Section IV, we have evaluated the constraints on the coupling parameter $b^2$ and the thermal conductivity $\lambda$ for the validity of GSLT using {\it Planck} data sets. A short discussion and concluding remarks have been presented in section V.

\section{A general prescription for the irreversible process}

In non-equilibrium thermodynamics, due to irreversiblity, there will be an internal entropy production. So in general, the change of entropy of a system can be written as
\begin{equation}
dS_{T}=d_{e}S+d_{i}S,
\end{equation}
where as before $d_{e}S$ stands for exchange of entropy between the system and its surroundings and $d_{i}S$ comes from internal production process. It should be noted that $d_{e}S$ may be positive, negative or zero depending upon the system's interaction with its surroundings, but $d_{i}S$ is always non-negative for an irreversible process.

If $\sigma$ and $\overrightarrow{J_{s}}$ stands for entropy production density and entropy flow density vector ({\it i.e.,} current) then under the assumption of local equilibrium [7-10],
\begin{eqnarray}
\frac{d_{e}S}{dt}&=&-\int_{\Sigma}\overrightarrow{J_{s}}d\overrightarrow{\Sigma}\nonumber\\
and~~~~\frac{d_{i}S}{dt}&=&\int_{V}\sigma dV,
\end{eqnarray}
where the volume $V$ is bounded by the surface $\Sigma$.

Now entropy flow may be caused by convection, heat conduction and diffusion, but we consider only heat conduction in order to have a simple physical picture. As a result we have
\begin{eqnarray}
\overrightarrow{ J_{s}} &=& \frac{\overrightarrow{ J_{q}}}{T}\nonumber \\
 and~~~\sigma &=&  \overrightarrow{ J_{q}}.\overrightarrow{\nabla}\left(\frac{1}{T}\right),
\end{eqnarray}
where $\overrightarrow{ J_{q}}$ stands for energy flux due to heat flow and $T$ is the temperature of the system. If we assume that energy flux and temperature remain constant across the surface $\Sigma$ then the first equation of (2) gives
\begin{equation}
\frac{d_{e}S}{dt}=4\pi R_{\Sigma}^{2}\frac{|\overrightarrow{J_{q}}|}{T}.
\end{equation}
However, if we assume Bekenstein's entropy area relation on the surface $\Sigma$ {\it i.e.,}
\begin{equation}
\frac{d_{e}S}{dt}= \frac{d}{dt}(\pi R_{\Sigma}^{2})=2\pi R_{\Sigma}\dot{R_{\Sigma}},
\end{equation}
then comparing (4) and (5) we have,
\begin{equation}
|\overrightarrow{J_{q}}|=\frac{T\dot{R_{\Sigma}}}{2R_{\Sigma}}.
\end{equation}
Similarly considering $\sigma$ to be uniform over the entire volume we have obtained from the second equation of (2) and using (3)
\begin{equation}
\frac{d_{i}S}{dt}=\frac{4}{3}\pi R_{\Sigma}^{3}\overrightarrow{J_{q}}.\overrightarrow{\nabla}\left(\frac{1}{T}\right).
\end{equation}

Now suppose we consider a heat flow in a non-accelerating, non-expanding and vorticity free fluid in flat space time and choose the comoving instaneous orthogonal frame as a global orthogonal frame. Applying energy conservation equation to the energy density ({\it i.e.,} $\rho=\frac{3}{2}nT$) for dust model based on the relativistic kinetic theory, we obtain
\begin{equation}
\overrightarrow{\nabla}\overrightarrow{ J_{q}}=\frac{3}{2}n\frac{\partial T}{\partial t},
\end{equation}
where the number density $n$ is assumed to be constant.

According to Eckart-Fourier law [11],
\begin{equation}
\overrightarrow{ J_{q}}=-\lambda\overrightarrow{\nabla}T,
\end{equation}
which states that there will be an energy flux if there is a temperature gradient ($\lambda$ is the thermal conductivity). 

Now combining (8) and (9), we have the usual heat conduction equation
\begin{equation}
\frac{\partial T}{\partial t}=\kappa \nabla^{2}T,
\end{equation}
with $\kappa=\frac{2\lambda}{3n}$. Note that due to parabolic nature of the above differential equation, there will be an infinite speed of propagation. Now eliminating $\overrightarrow{\nabla} T$ between equations (7) and (9) and using (6), we obtain
\begin{equation}
\frac{dS_{i}}{dt}=\frac{\pi R_{\Sigma}\dot{R}_{\Sigma}^{2}}{3\lambda}.
\end{equation}
Hence combining equations (5) and (11) the change of total entropy is given by
\begin{equation}
\frac{dS_{T}}{dt}=2\pi R_{\Sigma}\dot{R}_{\Sigma}\left(1+\frac{\dot{R}_{\Sigma}}{6\lambda}\right).
\end{equation}
In recent past non-equilibrium thermodynamics of FRW model of spacetime with the above modification has been studied [7,8,9] both at the apparent horizon and at the event horizon respectively for the dark energy fluid having constant or variable (holographic) equation of state. We have seen that the entropy variation due to production process is always positive irrespective of the sign of $\dot{R}_{\Sigma}$. 

However, the above Eckart theory has the following demerits namely (a) casuality violation, (b) describes unstable equilibrium states, (c) unable to describe the dynamics. Further, in a thermodynamical system it is expected that if a thermodynamical influence is switched off then the corresponding thermodynamic effect should be eliminated over a finite time period. But in the above Eckart theory, if temperature gradient is set to zero ({\it i.e.,} at $t=0$, $\vec{\nabla} T=0$) then from equation (9), $|\overrightarrow{J_{q}}|=0$ for $t\geq 0$, instead $|\overrightarrow{J_{q}}|$ gradually being zero after some finite period of time {\it i.e.,} expected form of $\overrightarrow{J_{q}}$ is
\begin{equation}
\overrightarrow{J_{q}}=\overrightarrow{J_{0}}exp\left(-t/\tau\right),
\end{equation}
where $\tau$ is a characteristic relaxation time for transient heat flow effects. Consequently the Fourier law is modified as [11,12]
\begin{equation}
\tau \dot{\overrightarrow{J_{q}}}+\overrightarrow{J_{q}}=-\lambda \overrightarrow{\nabla} T.
\end{equation}
This is known as Maxwell-Cattaneo modified Fourier law [11]. Now eliminating $\overrightarrow{J_{q}}$ between (8) and (14), we have the damped wave equation (instead of heat conduction equation)
\begin{equation}
\tau\frac{\partial^{2}T}{\partial t^{2}}+\frac{\partial T}{\partial t}-\lambda \nabla^{2}T=0.
\end{equation}
So for a thermal plane wave solution
\begin{equation}
T=T_{0}exp[i(\overrightarrow{k}.\overrightarrow{r}-w t)],
\end{equation}
the phase velocity is
\begin{equation}
V=\left[\frac{2\chi w}{\tau w+\sqrt{1+\tau^{2}w^{2}}}\right]^{1/2},
\end{equation}
and the dispersion relation takes the form [12,13]
\begin{equation}
|\kappa|^{2}=\frac{\tau w^{2}}{\chi}+iw.
\end{equation}
Thus in the high frequency limit ({\it i.e.,} $w\gg \tau^{-1}$), we have $V\simeq \sqrt{\frac{\chi}{\tau}}$ which is finite for $\tau >0$ and gives the speed of thermal pulses. Thus by introducing relaxation term it is possible to remove the problem of infinite propagation speed.

Now we shall determine the change of entropy due to internal production process by using modified Maxwell-Cattaneo Fourier law. As before, eliminating the temperature gradient term between equation (7) and (14) and using (6) for the magnitude of the energy flux we have,
\begin{equation}
\frac{d_{i}S}{dt} =\frac{\pi R_{\Sigma}\dot{R}_{\Sigma}^{2}}{3\lambda}\left[1+\frac{\tau \dot{u}}{2}\right],
\end{equation}
where $u=ln|\overrightarrow{J_{q}}|^{2}$, $|\overrightarrow{J_{q}}|=\frac{T\dot{R}_{\Sigma}}{2R_{\Sigma}}$.
So the total entropy change is
\begin{equation}
\frac{dS_{T}}{dt}=2\pi R_{\Sigma}\dot{R}_{\Sigma}\left[1+\frac{\dot{R}_{\Sigma}}{6\lambda}\left(1+\frac{\tau \dot{u}}{2}\right)\right].
\end{equation}

\section{Entropy Variation for Universe bounded by Apparent/Event horizon: Validity of GSLT}

In this section, we shall determine the time variation of total entropy of the universe bounded by the apparent/event horizon for flat FRW model and for different fluid distribution. If $R_{A}$ and $R_{E}$ denote the radius of the apparent and event horizon respectively then their time variation are given by
\begin{equation}
\dot{R}_{A}=-\frac{\dot{H}}{H^{2}}=1+q~~~and~~~\dot{R}_{E}= HR_{E}-1,
\end{equation}
where by definition $R_{A}=\frac{1}{H}$ and $R_{E}=a\int_{t}^{\infty}\frac{dt}{a}$ and $q$ is the usual deceleration parameter. Note that the improper integral in $R_{E}$ converges when strong energy condition is violated.

\subsection{Cosmic substratum as perfect fluid with constant equation of state}

Suppose $p=\omega\rho$ be the equation of state for the perfect fluid where $\omega$ ($<-\frac{1}{3}$) is constant. The cosmological solution is
\begin{center}
$a=a_{0}(t-t_{0})^{\frac{1}{\alpha}}~~,~~\rho=\rho_{0}a^{-2\alpha}$,
\end{center}
where
\begin{center}
$\alpha=\frac{3}{2}(1+\omega)~~,~~a_{0}=\left\{\frac{\sqrt{3\rho_{0}}}{2}(1+\omega)\right\}^{\frac{1}{\alpha}}$,
\end{center}
and $\rho_{0}$ is the constant of integration. The explicit form of the horizon radii are
\begin{center}
$R_{A}=\alpha t~~,~~R_{E}=\frac{\alpha t}{1-\alpha}$.
\end{center}
Thus total entropy variation for both the horizons are given by
\begin{equation}
\frac{dS_{T}^{A}}{dt}=2\pi\alpha^{2}t\left[1+\frac{\alpha}{6\lambda}\left(1-\frac{2\tau}{t}\right)\right],
\end{equation}
and
\begin{equation}
\frac{dS_{T}^{E}}{dt}=2\pi\left(\frac{\alpha}{1-\alpha}\right)^{2}t\left[1+\frac{\alpha}{6\lambda(1-\alpha)}\left(1-\frac{2\tau}{t}\right)\right].
\end{equation}
Thus for the validity of the generalized second law of thermodynamics (GSLT) we have,
\begin{description}
\item[$\frac{dS_{T}^{A}}{dt}>0$] If $-1<\omega<-1/3$ and $\tau<\frac{t}{2}\left(1+\frac{6\lambda}{\alpha}\right)$  or if  $\omega<-1$ and $\tau>\frac{t}{2}\left(1+\frac{6\lambda}{\alpha}\right)$.
\item[$\frac{dS_{T}^{E}}{dt}>0$] If $-1<\omega<-1/3$ and $\tau<\frac{t}{2}\left(1+\frac{6\lambda(1-\alpha)}{\alpha}\right)$ or if $\omega<-1$ and $\tau>\frac{t}{2}\left(1+\frac{6\lambda(1-\alpha)}{\alpha}\right)$.
\end{description}

\subsection{Cosmic Substratum as perfect fluid with variable equation of state}

For this general fluid the time variation of the total entropy has the explicit form
\begin{equation}
\frac{dS_{T}^{A}}{dt}=\frac{2\pi(1+q)}{H}\left[1+\frac{1+q}{6\lambda}-\frac{\tau H}{6\lambda}(r+3q+2)\right],
\end{equation}
and
\begin{equation}
\frac{dS_{T}^{E}}{dt}= 2\pi R_{E}(HR_{E}-1)\biggl[1+ \frac{(HR_{E}-1)}{6\lambda}-\frac{\tau H} {6\lambda}\left\lbrace(2HR_{E}-1)+q(3HR_{E}-2)\right\rbrace  \biggr].
\end{equation}
for the bounding horizon as apparent and event horizon respectively. Here $r=\frac{\dddot{a}}{aH^{3}}$ is the usual state finder parameter.

Now the explicit restrictions for the validity of GSLT are the following:\\
\begin{center} \textbf{Apparent Horizon} \end{center}
\begin{center}
\begin{tabular}{|c|c|}
\hline
\multicolumn{2}{|c|}{Quintessence era: $1+q>0$} \\
\hline
$r+3q+2<0$ &  No restriction on $~\tau$ \\
\hline
 $r+3q+2>0$  & $\tau H <min\left\lbrace\frac{1+q}{2(r+3q+2)},\frac{6\lambda}{2(r+3q+2)}\right\rbrace$ \\
\hline
\multicolumn{2}{|c|}{Phantom era: $1+q<0$} \\
\hline
$r+3q+2<0$ &   $\tau H <\frac{|1+q|-6\lambda}{2|r+3q+2|}$\\
\hline
 $r+3q+2>0$  & No restriction on $\tau$ \\
\hline
\end{tabular}
\end{center}
~~\\

\begin{center} \textbf{Event Horizon} \end{center}
\begin{center}
\begin{tabular}{|c|c|}
\hline
$|q|< \frac{2HR_{E}-1}{3HR_{E}-1}$ & $\tau H < \frac{HR_{E}-1}{2\left[(2HR_{E}-1)-|q|(2HR_{E}-1)\right]}$ \\
\hline
$|q|> \frac{2HR_{E}-1}{3HR_{E}-1}$ & $\tau$ is unrestricted \\
\hline 
\end{tabular}
\end{center}

\subsection{Cosmic fluid as interacting Dark species }

We consider interacting dark matter (DM) and holographic dark energy (HDE) as the matter in the  universe. The form of interaction term is chosen as [14,15]
\begin{equation}
Q=3b^{2}H(\rho_{m}+\rho_{D}),
\end{equation}
where the coupling parameter $b^{2}$ is assumed to be very small, $\rho_{m}$ is the energy density of the DM (in the form of dust) and $\rho_{D}$, the energy density for holographic dark energy satisfies (from holographic principle and effective field theory) [15,16]
\begin{equation}
\rho_{D}=\frac{3c^{2}M_{p}^{2}}{L^{2}},
\end{equation}
where $L$ is an IR cut-off in units $M_{p}^{2}=1$ and $c$ is any dimensionless parameter estimated by observations [17]. Here we choose radius of the event horizon as the IR cut-off length to obtain correct equation of state and the desired accelerating universe. So we have
\begin{equation}
R_{E}=\frac{c}{\sqrt{\Omega_{D}}H},
\end{equation}
where $\Omega_{D}=\frac{8\pi\rho_{D}}{3H^{2}}$ is the density parameter. Now the equation of state parameter of the holographic DE has the form [14,15,18] 
\begin{equation}
\omega_{D}=-\frac{1}{3}-\frac{2\sqrt{\Omega_{D}}}{3c}-\frac{b^{2}}{\Omega_{D}} 
\end{equation}
and the evolution of the density parameter is given by [14,15,18]
\begin{equation}
\Omega^{\prime}_{D}=\Omega_{D}^{2}(1-\Omega_{D})\left(\frac{1}{\Omega_{D}}+\frac{2}{c\sqrt{\Omega_{D}}}\right)-3b^{2}\Omega_{D},
\end{equation}
where '$\prime$' stands for the differentiation with respect to $x=lna$.

The Friedmann equations for the present interacting two fluid system are
\begin{equation}
H^{2}=\frac{8\pi G}{3}(\rho_{m}+\rho_{D})~~and~~\dot{H}=-4\pi G \left[\rho_{m}+(1+\omega_{D}\rho_{D})\right].
\end{equation}
Hence the deceleration parameter $q$ and the state finder parameter $r$ has the expressions
\begin{equation}
q=-\left(1+\frac{\dot{H}}{H^{2}}\right)=\frac{1}{2}\left(1-3b^{2}-\Omega_{D}-\frac{2\Omega_{D}^{3/2}}{c}\right),
\end{equation}
and
\begin{equation}
r=-2-3q-\frac{\Omega_{D}}{2}\left(1+\frac{2\sqrt{\Omega_{D}}}{c}\right)\left(\frac{3}{c}\sqrt{\Omega_{D}}-\frac{\Omega_{D}^{3}}{c}+6b^{2}-5\right)+\frac{9}{2}-\frac{3b^{2}}{2}\left\lbrace 3(2-b^{2})+\Omega_{D}\left(1+\frac{3}{c}\sqrt{\Omega_{D}}\right)\right\rbrace.
\end{equation}
Also the radius of the horizons and their time evolution are given by
\begin{equation}
R_{A}=\frac{1}{H}~~,~~R_{E}=\frac{c}{\sqrt{\Omega_{D}}H};
\end{equation}
and
\begin{equation}
\dot{R}_{A}=\frac{3}{2}\left(1-b^{2}-\frac{\Omega_{D}}{3}-\frac{2\Omega_{D}^{3/2}}{3c}\right)~~,~~\dot{R}_{E}=\frac{c}{\sqrt{\Omega_{D}}}-1.
\end{equation}
Hence the time variation of the total entropy of the universe bounded by apparent/event horizon are given by
\begin{equation}
\fontsize{1pt}{1.2}
\frac{dS_{T}^{A}}{dt}=\frac{\pi}{H}\lbrace 3(1-b^2)-\Omega _d(1+\frac{2\sqrt{\Omega _d}}{c})\rbrace [1 +\frac{1}{12\lambda}\lbrace 3(1-b^2)-\Omega _d(1+\frac{2\sqrt{\Omega _d}}{c})\rbrace -\frac{\tau H}{6\lambda}\lbrace 9(1-b^2-\frac{\Omega _d}{3}(1+\frac{2\sqrt{\Omega _d}}{c}))^2+\Omega _d(1+\frac{3\sqrt{\Omega _d}}{c})\lbrace (1-\Omega _d)(1+\frac{2\sqrt{\Omega _d}}{c})-3b^2 \rbrace \rbrace]
\end{equation}
and
\begin{equation}
\frac{dS_{T}^{E}}{dt}=\frac{2\pi c}{H\sqrt{\Omega_{d}}}\left(\frac{c}{\sqrt{\Omega_{d}}}-1\right)\left[1+\frac{1}{6\lambda}\left(\frac{c}{\sqrt{\Omega_{d}}}-1\right)
-\frac{\tau H}{6\lambda}\left\lbrace\left(2\frac{c}{\sqrt{\Omega_{d}}}-1\right)+q\left(3\frac{c}{\sqrt{\Omega_{d}}}-2\right)\right\rbrace\right].
\end{equation}

\section{Constraints on $b^2$ and $\lambda$ from {\it Planck} data sets for validity of GSLT}

In March 2013, based on the first 15.5 months of {\it Planck} investigations, the European Space Agency (ESA) and the {\it Planck} Collaboration publicly made available the CMB data along with a lot of scientific results [19]. Due to complicated expressions in (36) and (37) and a handful of observable parameters, we have used three {\it Planck} data sets [20] to evaluate some realistic bounds on the thermal conductivity $\lambda$ and the coupling parameter $b^2$ (for arbitrary values of the relaxation time $\tau$) which make $\frac{dS_{T}^{A}}{dt}$ and $\frac{dS_{T}^{E}}{dt}$ non-negative. The non-negativity of the two quantities is necessary for GSLT to hold in both the cases.

Compared to WMAP results, the {\it Planck} results reduce the error by 30\% to 60\% and thus improves the constraints on dark energy. The results have been seen to differ significantly if the {\it Planck} data are combined with external astrophysical data sets such as the BAO measurements (can provide effective constraints on dark energy from the angular diameter distance$-$redshift relation) from 6dFGS+SDSS DR7(R)+BOSS DR9, the direct measurement of the Hubble constant, $H_0=73.8 \pm 2.4 km s^{-1} Mpc^{-1}$ (1$\sigma$ CL) [21], from the supernova magnitude$-$redshift relation calibrated by the HST observations of Cepheid variables in the host galaxies of eight SNe Ia and the supernova data sets: The SNLS3 (which is a "combined" sample [22], consisting of 472 SNe, calibrated by both SiFTO [23] and SALT2 [24]) and the Union 2.1 compilation [25], consisting of 580 SNe, calibrated by the SALT2 light-curve fitting model [24]. These external data sets contribute significantly to the accuracy of the constraint results. The lensing data further improves the constraints by 2\% to 15\%. Also, no tension has been found [20] when {\it Planck} data is combined with the BAO, HST and Union 2.1 data sets. However, combination of SNLS3 with other data sets shows a weak tension. Table I below shows the three {\it Planck} data sets and the observed values of the parameters which we shall use in order to compute the bounds on $\lambda$ and $b^2$ for arbitrary values of $\tau$. Recently, using {\it Planck} data, constraints on the coupling parameter $b^2$ have been evaluated [26] for validity of thermodynamical equilibrium in case of equilibrium thermodynamics.

\begin{center} {\bf Table-I}: {\it Planck} data sets \end{center} 
\begin{center}
\begin{tabular}{|c|c|c|}
\hline Data & $c$ & ${\Omega}_{d}$\\
\hline \hline {\it Planck}+WP+SNLS3+lensing & 0.603 & 0.699\\
\hline {\it Planck}+WP+BAO+HST+lensing & 0.495 & 0.745\\
\hline {\it Planck}+WP+Union 2.1+BAO+HST+lensing & 0.577 & 0.719\\

\hline
\end{tabular}
\end{center}

\vspace{0.4cm}

In the above table, "{\it Planck}" represents the {\it Planck} temperature likelihood [20] (including both the low-$l$ and high-$l$ parts), "WP" represents the WMAP polarization likelihood as a supplement of {\it Planck}, and "lensing" represents the likelihood of {\it Planck} lensing data, in reference to the likelihood software provided by the {\it Planck} Collaboration.

Table II shows the constraints on $b^2$ and $\lambda$ ($\tau$ can take arbitrary values) which are required for GSLT to hold in case of Universe bounded by apparent horizon.

\begin{center} {\bf Table-II}: Constraints on $b^2$ and $\lambda$ for GSLT to hold in case of apparent horizon \end{center} 
\begin{center}
\begin{tabular}{|c|c|c|c|}
\hline $c$ & $\Omega _d$ & $b^2$ & $\lambda$\\
\hline \hline 0.603 & 0.699 & 0.9490 $\leq$ $b^2$ $\leq$ 1  & $\lambda$ $\leq$ 0.2070\\
\hline 0.495 & 0.745 & All values of $b^2$ & $\lambda$ $\leq$ 0.0286\\
\hline 0.577 & 0.719 & 0.8236 $\leq$ $b^2$ $\leq$ 1 & $\lambda$ $\leq$ 0.1920\\
\hline
\end{tabular}
\end{center}

\begin{figure}
\begin{minipage}{0.4\textwidth}
\includegraphics[width=1.0\linewidth]{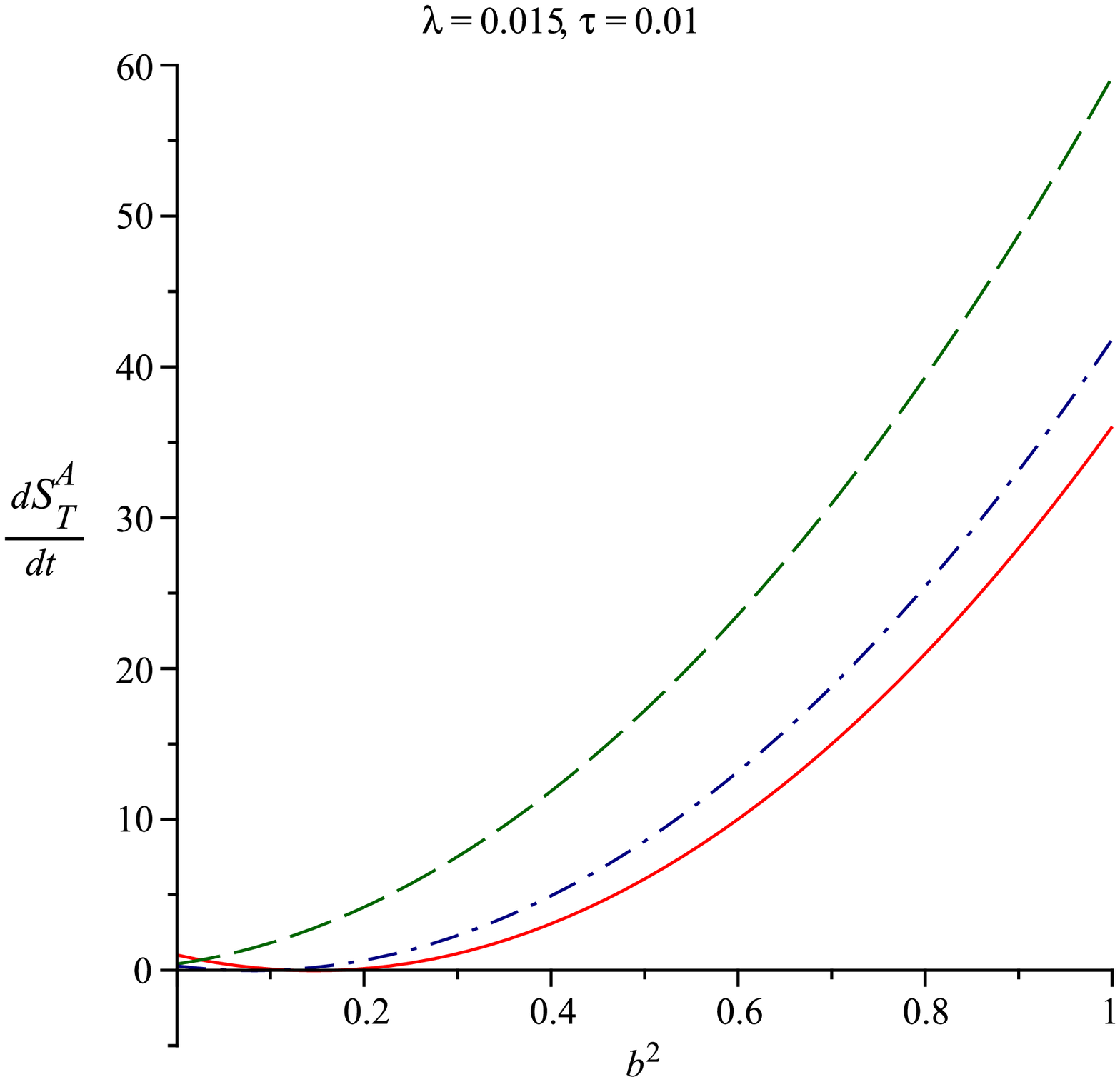}
\end{minipage}
\begin{minipage}{0.4\textwidth}
\includegraphics[width=1.0\linewidth]{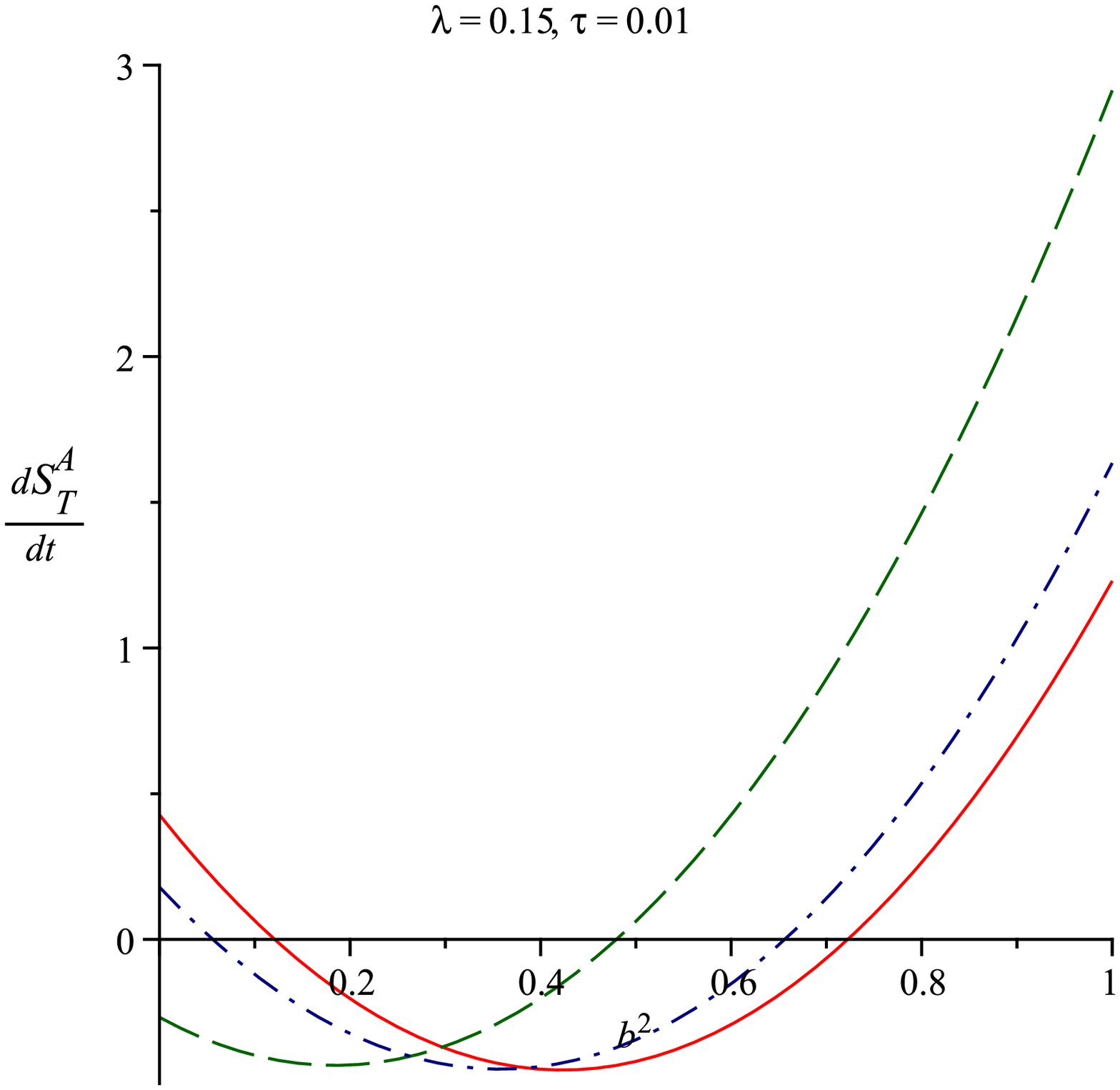}
\end{minipage}
\begin{minipage}{0.4\textwidth}
\includegraphics[width=1.0\linewidth]{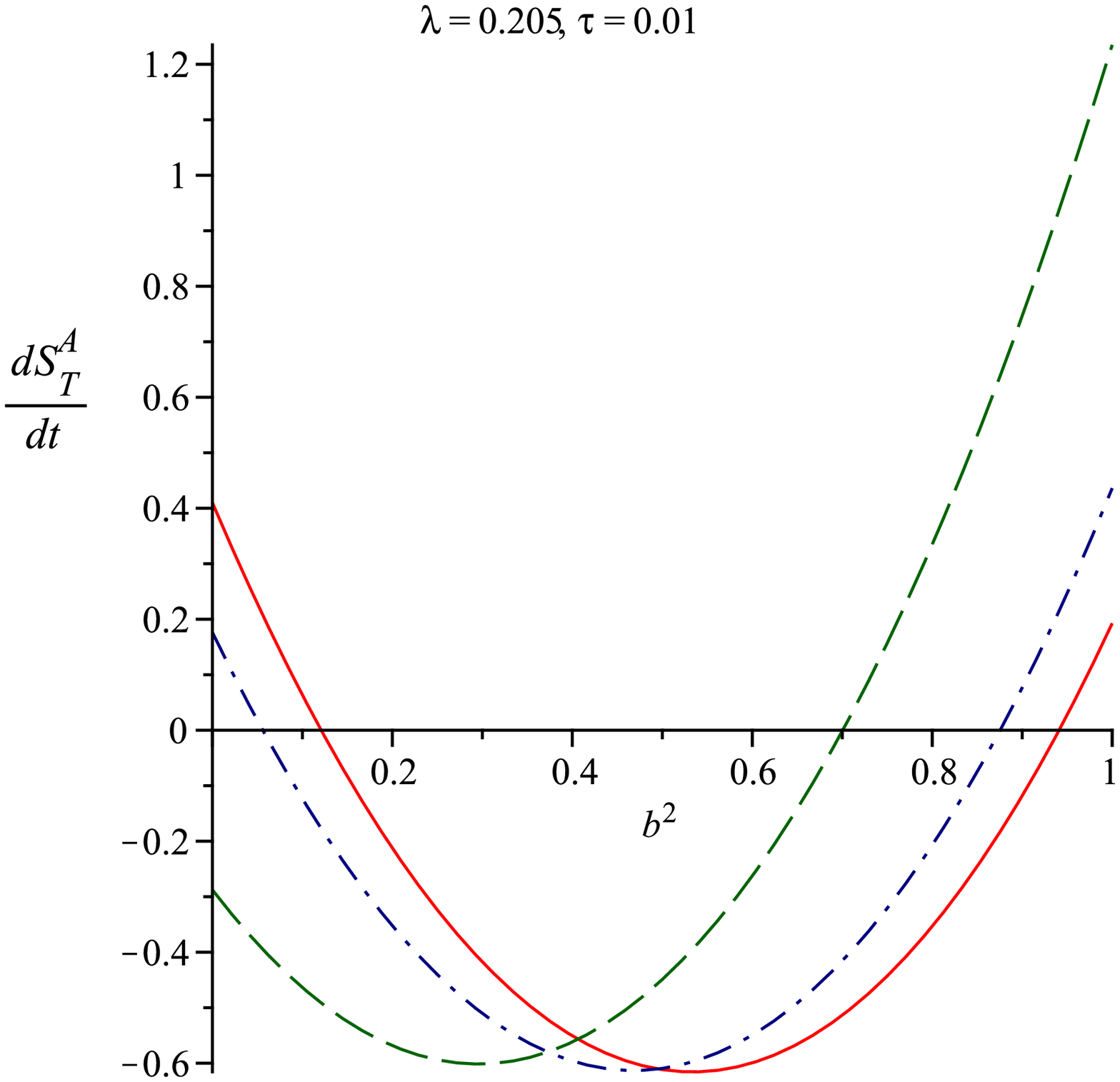}
\end{minipage}
\begin{minipage}{0.4\textwidth}
\includegraphics[width=1.0\linewidth]{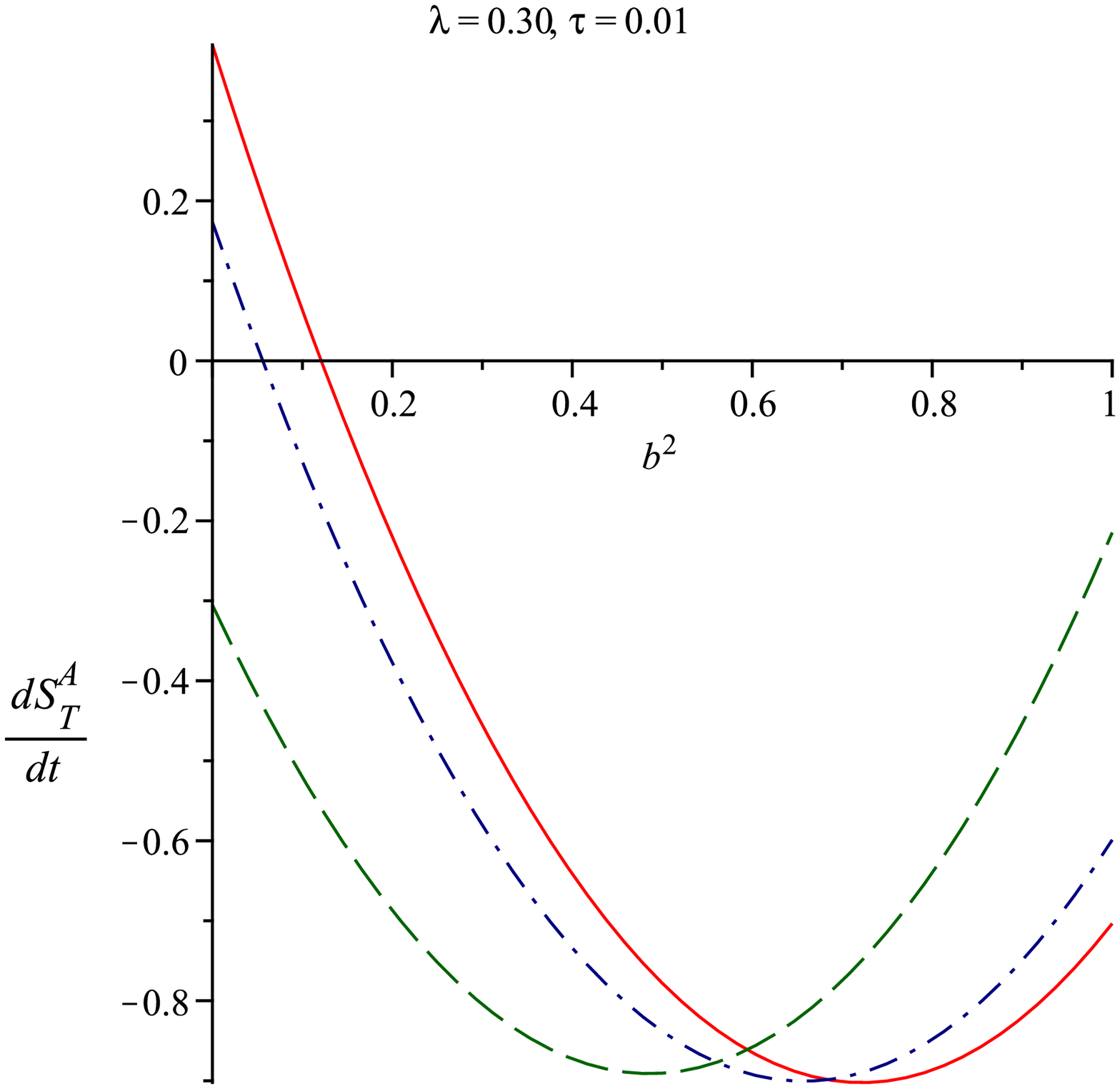}
\end{minipage}
\hspace{4cm}\\
\fontsize{5pt}{5.0}
Figure 1: The above 4 plots show the variation of $\frac{dS_{T}^{A}}{dt}$ against the coupling parameter $b^2$ for different values of $\lambda$ and for $\tau =0.01$. The red (solid), green (dash) and blue (dash-dot) curves correspond to data set 1, 2 and 3 respectively.
\end{figure}

\vspace{0.4cm}

Some remarks about the bounds obtained in Table II are in order. Let us consider the following three expressions ({\it see} Eq. (36)):
\begin{center}
$A=3(1-b^2)-\Omega _d\left(1+\frac{2\sqrt{\Omega _d}}{c}\right)$
\end{center}
\begin{center}
$B=1+\frac{A}{12\lambda}$
\end{center}
\begin{center}
$C=\left[ 9\left(1-b^2-\frac{\Omega _d}{3}\left(1+\frac{2\sqrt{\Omega _d}}{c}\right)\right)^2+\Omega _d\left(1+\frac{3\sqrt{\Omega _d}}{c}\right)\left\lbrace (1-\Omega _d)\left(1+\frac{2\sqrt{\Omega _d}}{c}\right)-3b^2 \right\rbrace \right]$
\end{center}
Our first task is to determine the signs of $A$, $B$ and $C$ which are required for GSLT to hold. One can easily note that $\frac{dS_{T}^{A}}{dt}$ is non-negative if ($A$,$B$,$C$) either has the sign combination (+,+,$-$) or ($-$,$-$,+). Since $c$ and $\Omega _d$ are observable parameters, expressions $A$ and $C$ together give the bounds on $b^2$. These bounds in turn give the bounds on $\lambda$ from the expression $B$. Now, for all the three data sets, there exists no value of $b^2$ for which $A$ becomes positive and $C$ becomes negative simultaneously. For instance, consider the data set 1 ({\it Planck}+WP+SNLS3+lensing). In this case, the expression $A$ is positive only if $b^2 \in (0,0.1209)$ while the expression $C$ is negative only if $b^2 \in (0.4949,0.9490)$. So, a common value (or a common range of values) for $b^2$ satisfying ($A$,$C$) $\rightarrow$ (+,$-$) is not possible to achieve. Thus, only the combination ($-$,$-$,+) is viable with which one can compute the bounds listed in Table II. Further, we have plotted $\frac{dS_{T}^{A}}{dt}$ against the coupling parameter $b^2$ for four different values of the thermal conductivity $\lambda$ and for $\tau =0.01$ (since the constraints obtained are independent of $\tau$). The values of $\lambda$ have been chosen such that the constraints obtained in Table II are reflected through these plots. For example, for $\lambda=0.205$ ({\it see} Plot 3 of Fig. 1), the constraints on $b^2$ (for GSLT to hold) are consistent with those obtained in Table II for data set 1 but not for data sets 2 or 3. 

We now turn our attention to the case of Universe bounded by event horizon. Table III shows the bounds on $b^2$ and $\lambda$ which make $\frac{dS_{T}^{E}}{dt}$ non-negative, $\tau$ being arbitrary.

\begin{figure}
\begin{minipage}{0.4\textwidth}
\includegraphics[width=1.0\linewidth]{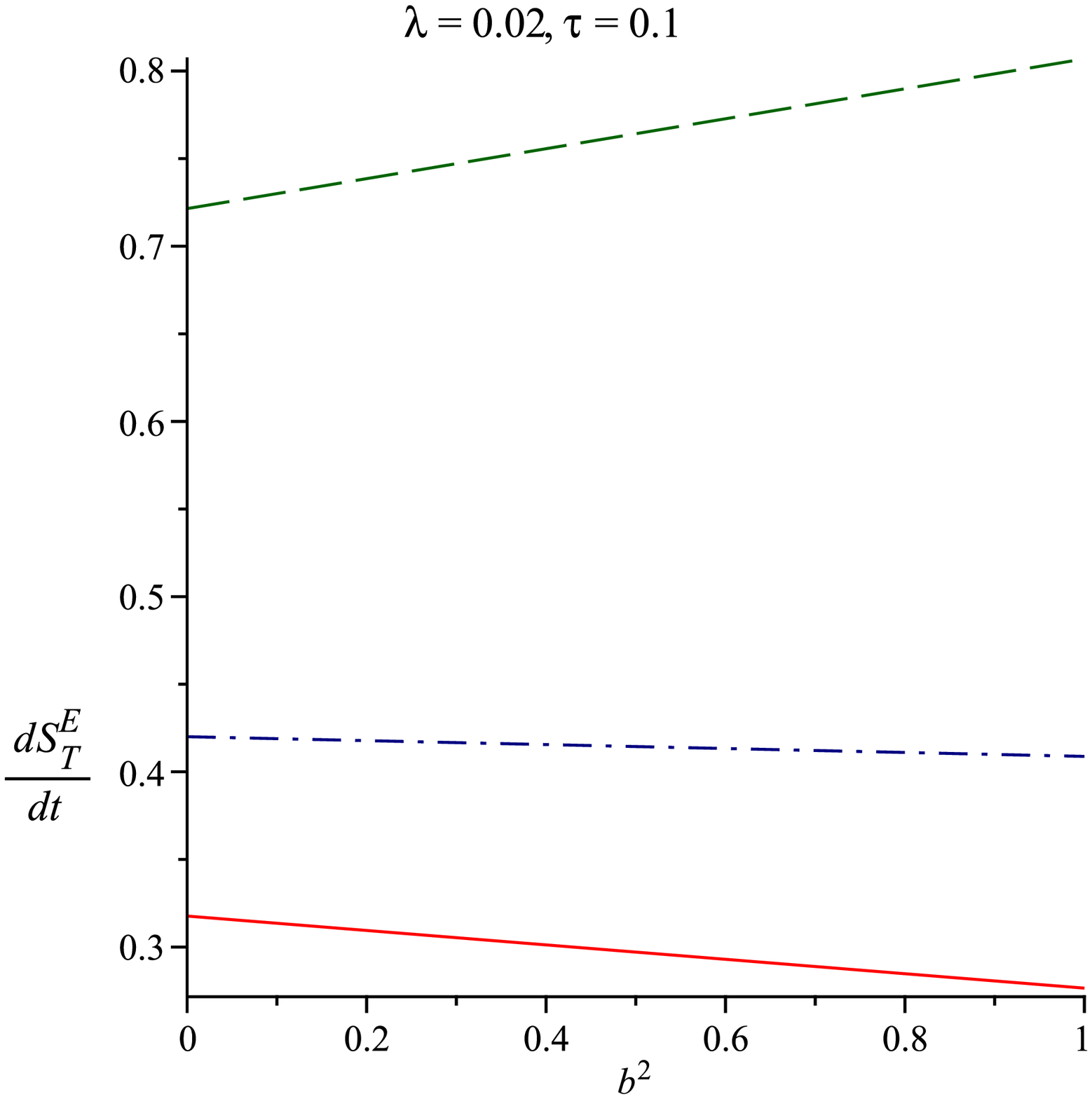}
\end{minipage}
\begin{minipage}{0.4\textwidth}
\includegraphics[width=1.0\linewidth]{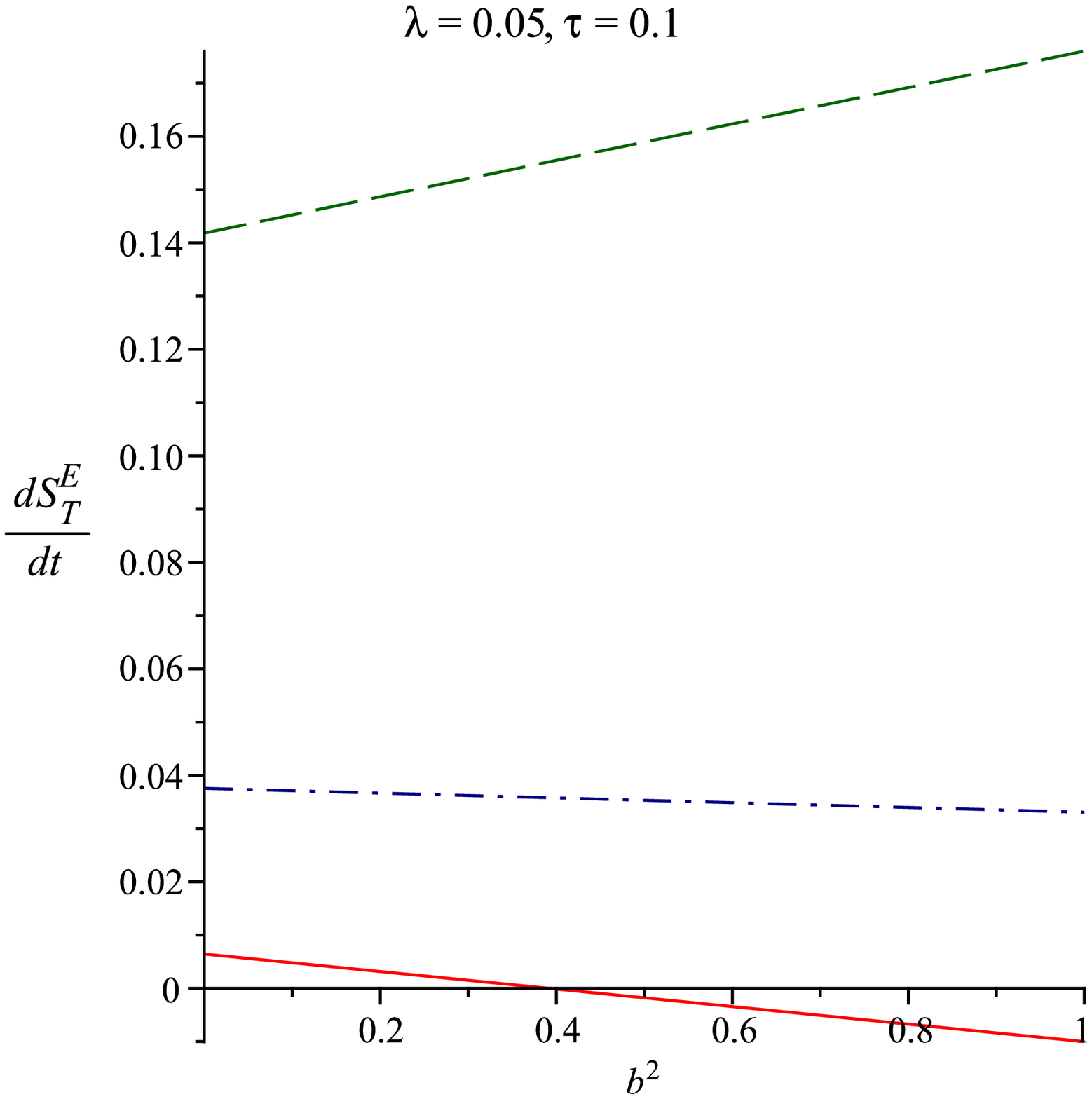}
\end{minipage}
\begin{minipage}{0.4\textwidth}
\includegraphics[width=1.0\linewidth]{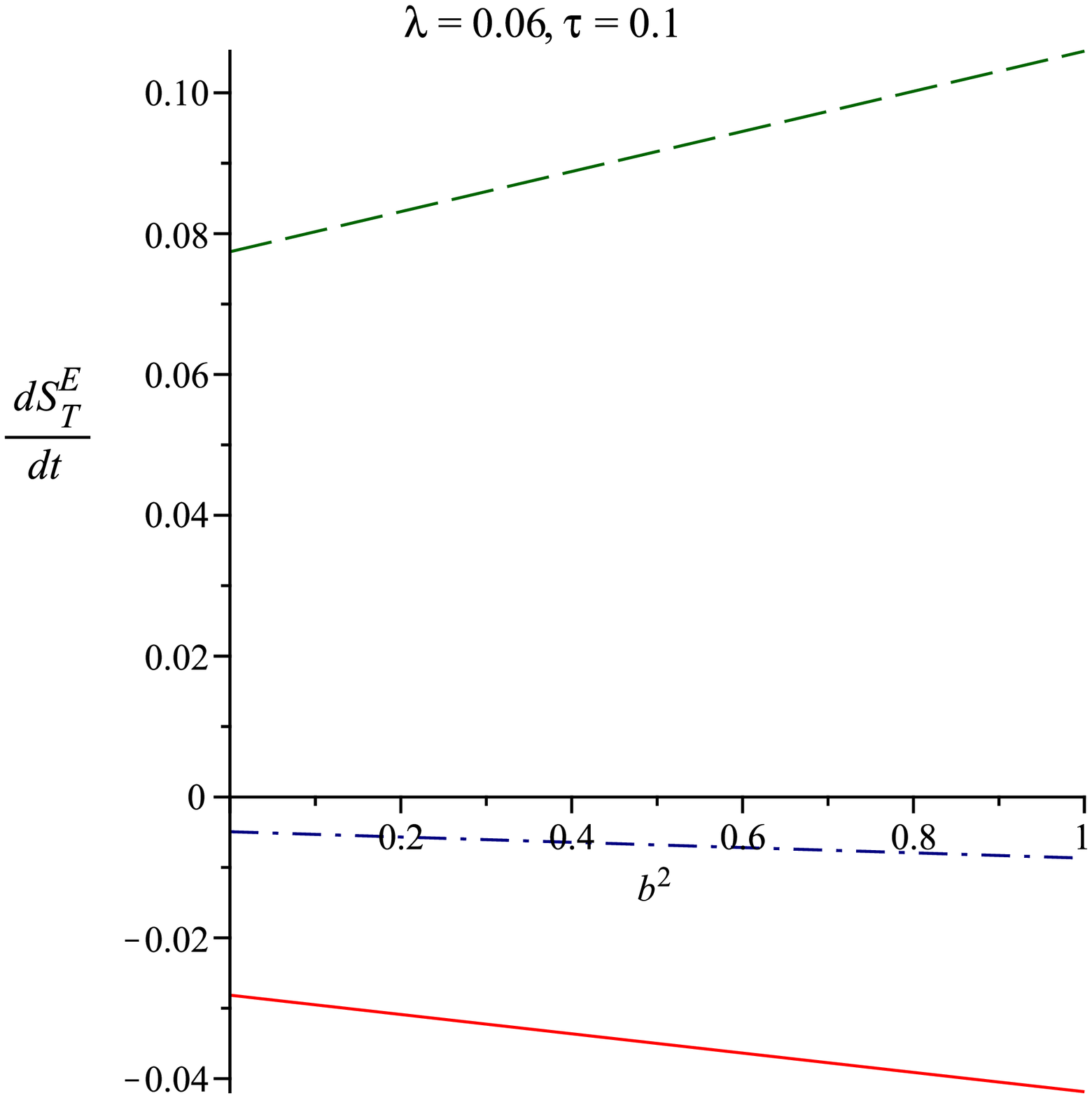}
\end{minipage}
\begin{minipage}{0.4\textwidth}
\includegraphics[width=1.0\linewidth]{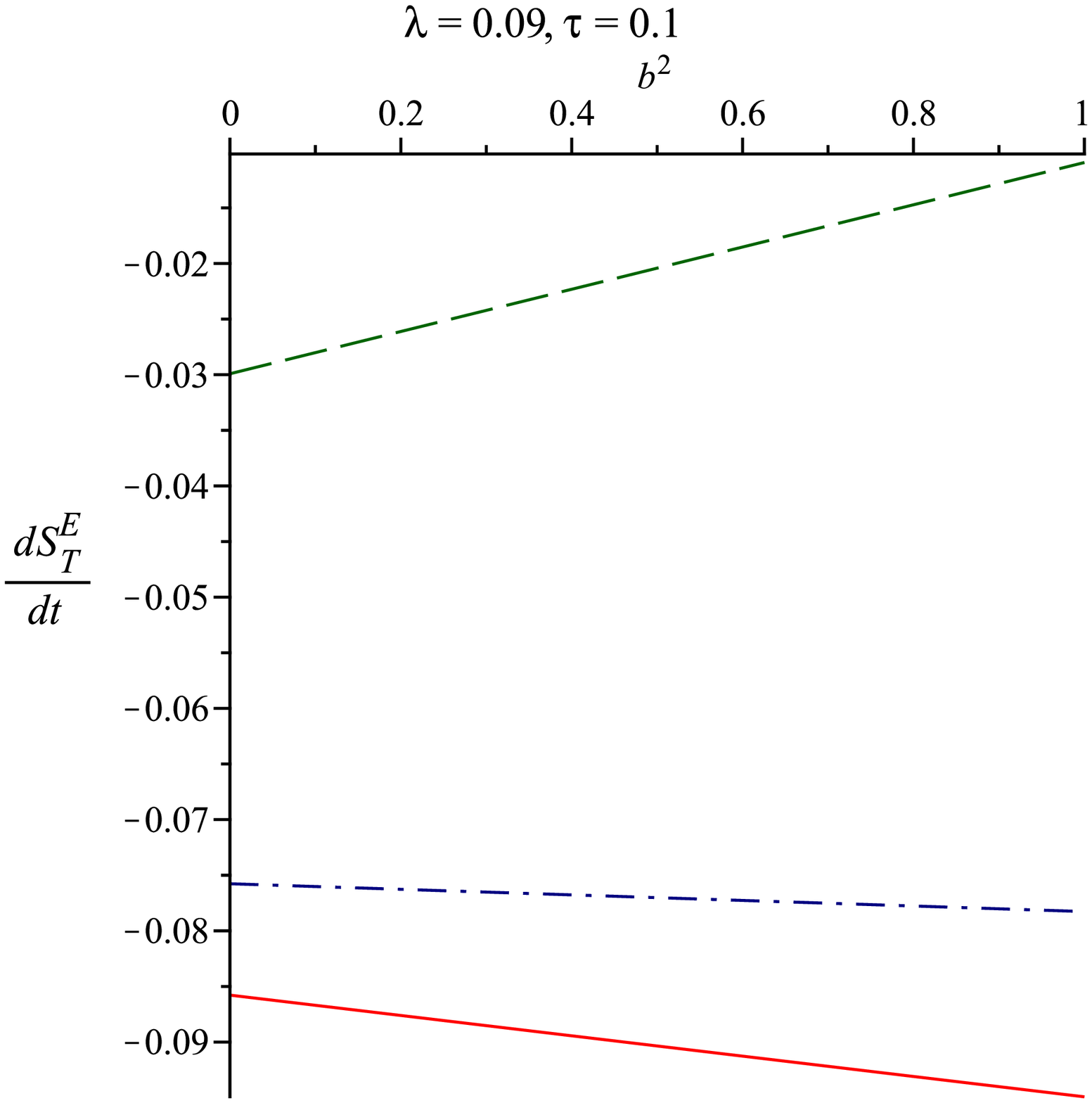}
\end{minipage}
\hspace{4cm}\\
\fontsize{5pt}{5.0}
Figure 2: The above 4 plots show the variation of $\frac{dS_{T}^{E}}{dt}$ against the coupling parameter $b^2$ for different values of $\lambda$ and for $\tau =0.1$. The red (solid), green (dash) and blue (dash-dot) curves correspond to data set 1, 2 and 3 respectively.
\end{figure}

\vspace{0.4cm}

\begin{center} {\bf Table-III}: Constraints on $b^2$ and $\lambda$ for GSLT to hold in case of event horizon \end{center} 
\begin{center}
\begin{tabular}{|c|c|c|c|}
\hline $c$ & $\Omega _d$ & $b^2$ & $\lambda$\\
\hline \hline 0.603 & 0.699 & All values of $b^2$  & $\lambda$ $\leq$ 0.0465\\
\hline 0.495 & 0.745 & All values of $b^2$ & $\lambda$ $\leq$ 0.0711\\
\hline 0.577 & 0.719 & All values of $b^2$ & $\lambda$ $\leq$ 0.0533\\
\hline
\end{tabular}
\end{center}

As in the case of apparent horizon, we consider the following expressions ({\it see} Eq. (37)):
\begin{center}
$D=\frac{c}{\sqrt{\Omega _d}}-1$
\end{center}
\begin{center}
$E=1+\frac{D}{6\lambda}$
\end{center}
\begin{center}
$F=\left \lbrace \left(2\frac{c}{\sqrt{\Omega_{d}}}-1\right)+\frac{1}{2}\left(1-3b^{2}-\Omega_{D}-\frac{2\Omega_{D}^{3/2}}{c}\right)\left(3\frac{c}{\sqrt{\Omega_{d}}}-2\right) \right \rbrace$
\end{center}
Here one can see that for all the three {\it Planck} data sets, the expression $D$ is negative. So, for evaluating the bounds, only the sign combination ($D$,$E$,$F$) $\rightarrow$ ($-$,$-$,+) must be considered. Here also, we have plotted $\frac{dS_{T}^{E}}{dt}$ against the coupling parameter $b^2$ for four different values of the thermal conductivity $\lambda$ and for $\tau =0.1$ in Fig. 2. One can easily see that these plots are consistent with the constraints obtained in Table III.

Moreover, one should also note that in Tables II and III, the constraints have been kept correct to 4 decimal places and the coupling parameter $b^2 \in [0,1]$. 

\section{Short Discussion and Conclusions}

An extensive study of the irreversible thermodynamics of the Universe is considered for flat FRW model. For simplicity, we consider the entropy flow only due to heat conduction. To incorporate relaxation time, the Fourier law is modified as Maxwell-Cattaneo modified Fourier law and as a result the usual heat conduction equation changes to damped wave equation.

Subsequently, generalized second law of thermodynamics is examined for three choices of the cosmic fluid namely perfect fluid with constant or variable equation of state and interacting holographic dark energy and dark matter. For validity of GSLT, analytic inequalities are possible for perfect fluid (both constant and variable equation of state) while for holographic dark energy model, the expressions for total entropy variation with time has complicated expressions for both the horizons (apparent and event). So using {\it Planck} data sets for the observed values of the dimensionless parameter $c$ and the density parameter $\Omega _d$, we have estimated the admissible range of the coupling parameter $b^2$ and the thermal conductivity $\lambda$ ($\tau$ being arbitrary) for the validity of GSLT. Also, graphically we have shown the variation of time variation of total entropy against $b^2$ for allowed choices of $\lambda$ and for arbitrary $\tau$. From Tables II and III, we see that for three data sets GSLT holds for all values of $b^2$ for event horizon while for apparent horizon, $b^2$ is unrestricted only for one data set and this is reflected in Figs. 1 and 2. Lastly, we note that in equilibrium thermodynamics, there is no restriction for validity of GSLT across apparent horizon (for any gravity theory) while there are some realistic conditions for validity of GSLT bounded by event horizon but GSLT holds in a restrictive way for both the horizons in irreversible thermodynamics. Therefore, based on the present work we may conclude that in non-equilibrium prescription of thermodynamics, event horizon is more favourable than apparent horizon for FRW model of the Universe.\\

$~~~~~~~~~~~~~~~~~~~~~~~~~~~~~~~~~~~~~~~~~$\textbf{ACKNOWLEDGEMENTS}\\

The author S.S. is thankful to UGC-BSR Programme of Jadavpur University for awarding Research Fellowship. Author S.C. is thankful to UGC-DRS Programme in the Department of Mathematics, Jadavpur University.



\begin{thebibliography}{26}
\bibitem{Hawking} S.W. Hawking,{ \it Commun. Math. Phys.} {\bf 43}, 199 (1975).
\bibitem{Bekenstein} J.D. Bekenstein, {\it Phys. Rev. D} {\bf 7}, 2333 (1973).
\bibitem{Bardeen} J.M. Bardeen, B. Carter, S.W. Hawking, {\it Commun. Math. Phys.} {\bf 31} 161 (1973).
\bibitem{Jacobson} T. Jacobson, {\it Phys. Rev. Lett.} {\bf 75}, 1260 (1995).
\bibitem{Padmanabhan} T. Padmanabhan, {\it Class. Quantum Grav.} {\bf 19}, 5387 (2002).
\bibitem{Guedens} C. Eling, R. Guedens, T. Jacobson, {\it Phys. Rev. Lett.} {\bf 96}, 12301(2006); C. Eling, {\it JHEP} {\bf 11}, 048(2008).
\bibitem{Gang} W. Gang, L. Wen-Biao, {\it Commun. Theor. Phys.} {\bf 52}, 383 (2009).
\bibitem{Chakraborty1} S. Chakraborty, A. Biswas, {\it Astrophys. Space. Sci.} {\bf 343}, 395 (2013).
\bibitem{Chakraborty2} S. Chakraborty, A. Biswas, {\it Astrophys. Space Sci.} {\bf 343}, 791 (2013).
\bibitem{Karami1} K. Karami, M. Jamil, N. Sahraei, {\it Phys. Scripta} {\bf 82}, 045901 (2010).
\bibitem{Maartens} R. Maartens, arXiv: astro-ph/9609119.
\bibitem{Zimdahl1} W. Zimdahl, {\it Phys. Rev. D} {\bf 61}, 083511 (2000).
\bibitem{Zimdahl2} W. Zimdahl, {\it Phys. Rev. D} {\bf 53}, 5483 (1996).
\bibitem{Wang1} B. Wang, Y.G. Gong, and E. Abdalla, {\it Phys. Rev. D} {\bf 74}, 083520 (2006).
\bibitem{Li} M. Li, {\it Phys. Lett. B} {\bf 603}, 1 (2004).
\bibitem{Mazumder1} N. Mazumder and S. Chakraborty, {\it Gen. Rel. Grav.} {\bf 42} 813 (2010).
\bibitem{Cohen} A. Cohen, D. Kaplan, A. Nelson, {\it Phys. Rev. Lett.} {\bf 82} 4971 (1999).
\bibitem{Wang2} B. Wang, Y.G. Gong and E. Abdalla, {\it Phys. Lett. B} {\bf 624}, 141 (2005).
\bibitem{Ade} P.A.R. Ade {\it et al.} {\it Planck} Collaboration, arXiv: 1303.5062 [astro-ph.CO].
\bibitem{Li} M. Li, X.D. Li, Y.Z. Ma, X. Zhang and Z. Zhang, {\it JCAP} {\bf 09}, 021 (2013).
\bibitem{Riess} A. G. Riess {\it et al.}, {\it Astrophys. J.} {\bf 730}, 119 (2011).
\bibitem{Guy1} J. Guy {\it et al.}, {\it Astron. Astrophys.} {\bf 523}, A7 (2010); M. Sullivan {\it et al.}, {\it Astrophys. J.} {\bf 737}, 102 (2011).
\bibitem{Conley} S. Conley {\it et al.}, {\it Mon. Not. Roy. Astron. Soc.} {\bf 362}, 505 (2008).
\bibitem{Guy2} J. Guy {\it et al.}, {\it Astron. Astrophys.} {\bf 466}, 11 (2007).
\bibitem{Suzuki} N. Suzuki {\it et al.}, {\it Astrophys. J.} {\bf 746}, 85 (2012).
\bibitem{Saha1} S. Saha and S. Chakraborty, {\it Phys. Rev. D} {\bf 89}, 043512 (2014).

\end{thebibliography}
\end{document}